\documentclass[reprint,amsmath,amssymb,aps,prd,superscriptaddress]{revtex4-1}

\usepackage{mathtools}
\usepackage{float}
\usepackage{graphicx}
\usepackage{dcolumn}
\usepackage{bm}
\usepackage{subcaption} 
\usepackage{hyperref}
\usepackage{url}
\usepackage{physics}

\usepackage{soul}

\usepackage[usenames,dvipsnames]{xcolor}
\colorlet{BLUE}{blue}
\colorlet{RED}{red}

\def\Weizmann{\small{Department of Particle Physics and Astrophysics, Weizmann Institute of Science, Rehovot 761001, Israel}}
\def\JGU{\small{Johannes Gutenberg-Universit{\"a}t Mainz, Helmholtz-Institut Mainz, GSI Helmholtzzentrum f{\"u}r Schwerionenforschung, 55128 Mainz, Germany}}
\def\HHU{\small{Heinrich-Heine-Universit{\"a}t D{\"u}sseldorf, 40225 D{\"u}sseldorf, Germany}}
\def\UCB{\small{Department of Physics, University of California, Berkeley, California 94720, USA}}

\usepackage{tikz,xcolor,hyperref}
\definecolor{lime}{HTML}{A6CE39}
\DeclareRobustCommand{\orcidicon}{%
	\begin{tikzpicture}
	\draw[lime, fill=lime] (0,0) 
	circle [radius=0.16] 
	node[white] {{\fontfamily{qag}\selectfont \tiny ID}};	\draw[white, fill=white] (-0.0625,0.095) 
	circle [radius=0.007];	\end{tikzpicture}
	\hspace{-2mm}}
\foreach \x in {A,...,Z}{%
	\expandafter\xdef\csname orcid\x\endcsname{\noexpand\href{https://orcid.org/\csname orcidauthor\x\endcsname}{\noexpand\orcidicon}}
}

\begin{document}

\title{Search for ultralight dark matter with spectroscopy of radio-frequency atomic transitions}





%
%

\author{Xue Zhang\orcidA{}}
\affiliation{\JGU}

\author{Abhishek Banerjee\orcidB{}}
\affiliation{\Weizmann}

\author{Mahapan Leyser
}
\affiliation{\JGU}

\author{\mbox{Gilad Perez\orcidD{}}}
\affiliation{\Weizmann}

\author{Stephan Schiller\orcidE{}}
\affiliation{\HHU}

\author{Dmitry Budker\orcidF{}}
\affiliation{\JGU}
\affiliation{\UCB}

\author{Dionysios Antypas\orcidG{}\,}
\email{dantypas@uni-mainz.de}
\affiliation{\JGU}


\date{\today}

\begin{abstract}
 

The effects of scalar and pseudoscalar ultralight bosonic dark matter (UBDM) were searched for by 
comparing the frequency of a quartz oscillator to that of a 
hyperfine-structure transition in $^{87}$Rb, and an
electronic transition in $^{164}$Dy. We constrain linear interactions between a scalar UBDM field and Standard-Model (SM) fields for an underlying UBDM particle mass in the range 
\mbox{$1\times10^{-17}-8.3\times10^{-13} $\,eV} and quadratic interactions between a pseudoscalar UBDM field and SM fields in the range \mbox{$5\times10^{-18}- 4.1\times10^{-13} $\,eV.}
Within regions of the respective ranges, our constraints on linear interactions significantly improve on results from previous, direct searches for oscillations in atomic parameters, while constraints on quadratic interactions surpass limits imposed by such direct searches as well as by astrophysical observations.
 
 \end{abstract}

\maketitle

\emph{Introduction---} Apparently, dark matter (DM) makes up the majority of matter in our Universe 
\cite{WorkmanPDG2022}, as indicated by decades of astronomical and cosmological observations \cite{KimballBook}, and yet the nature and composition of DM remain unknown.
There is a broad class of well-motivated models, where the 
DM constituent is a spin-0 particle with mass in the range of $m_{\phi}\approx10^{-22}-$ 10\,eV \cite{KimballBook,SafronovaRMP2018}. These ultralight bosonic dark matter (UBDM) 
particles are predicted 
to behave locally like a classical 
field,
coherently oscillating at the particle's Compton frequency $f_{\rm C}= m_{\phi}/(2\pi)$. 


The interaction between the UBDM field and the Standard-Model (SM) 
fields varies between different models, according to the UBDM symmetry properties such as CP 
(see~\cite{Banerjee:2022sqg} for a recent discussion). 
The UBDM particle could be a parity-even scalar field 
(dilaton), associated with spontaneous breaking of the scale-invariance symmetry (see for example~\cite{ArvanitakiPRD2015,GrahamARNPC2015b}). Alternatively, it could be a
relaxion, a special kind of axion-like particle,  that couples to SM matter, dominantly, via its mixing with the Higgs boson~\cite{GrahamPRL2015,BanerjeePRD2019,FlackeJHEP2017,Banerjee:2020kww}. 
Interestingly, even in the celebrated case of the CP-odd QCD axion~\cite{Preskill:1982cy,Abbott:1982af,Dine:1982ah}, originally proposed to explain the smallness of CP-violation in the strong force~\cite{Peccei:1977ur,Peccei:1977hh,Weinberg:1977ma,Wilczek:1977pj,Kim:1979if,Shifman:1979if,Zhitnitsky:1980tq,Dine:1981rt}, there are quadratic-scalar interactions between the QCD-axion field and the SM ones~\cite{Kim:2022ype}. Further exotic models dominated by a quadratically-coupled UBDM were recently described in~\cite{Banerjee:2022sqg}. 


An interaction between an ultralight scalar field and SM fields may induce violation of Einstein's equivalence principle (EP)\cite{Damour:2010rp,DamourPRD2010, HeesPRD2018} 
and oscillations in the fundamental constants (FCs) 
of nature 
\cite{ArvanitakiPRD2015,BanerjeePRD2019}.
Such FCs include the fine-structure constant $\alpha$, electron mass $m_{\rm e}$, and constants which determine the nuclear mass, for instance, the QCD energy scale $\Lambda_{\rm{QCD}}$ and the quark masses. 
For a QCD-axion UBDM,  oscillating nucleon electric-dipole moments are expected in case of linear coupling, \cite{Graham:2013gfa}, and, as pointed out recently \cite{Kim:2022ype}, 
oscillations in nuclear parameters such as nucleon masses and nuclear $g$-factors are also predicted due to the presence of quadratic coupling (see also \cite{StadnikFlambaumPRL2015} for discussion on oscillating FCs due to quadratic coupling to the SM fields). 


Experiments designed to check for EP violation offer a way to probe scalar UBDM \cite{SmithPRD1999,SchlammingerPRL2008,TouboulPRL2017,BergePRL2018}. 
Other works aim to detect the effects of light scalar fields by searching for oscillations in FCs. These would appear as oscillations in the length or density of solids, or in the energies of atomic or molecular levels. Various searches were proposed or completed  \cite{ArvanitakiPRL2016,ManleyPRL2020,GeraciPRL2019,StadnikPRL2015, StadnikPRA2016, GrotePhysRevResearch2019,SavallePRL2020, Vermeulennature2021,AielloPRL2022,VanTilburgPRL2015, HeesPRL2016,WiczloSciAdv2018,BeloyNature2021,KennedyPRL2020,AharonyPRD2021,CampbellPRL2020,AntypasPRL2019,Wrestle2, FlambaumArxiv2022,HannekeQST2020, AntypasQST2021, OswaldPRL2021}; see \cite{AntypasSnowmass2022} for a review of experimental activities.  Pseudoscalar UBDM may also introduce oscillatory effects in atomic or molecular systems \cite{Kim:2022ype,StadnikFlambaumPRL2015}, 
yielding observables that are indistinguishable from these due to scalar UBDM. This enables one to probe both classes of UBDM models with the same apparatus.

Here we search for the effects of scalar UBDM 
in two distinct experiments, where we compare the frequency of a quartz oscillator to the frequency of either of two radio-frequency (rf) transitions: a hyperfine transition between electronic ground levels in ${}^{87}$Rb (Experiment 1), and an electric-dipole transition between two nearly degenerate states in $^{164}$Dy (Experiment 2). These searches are implemented in the UBDM particle mass range 
\mbox{$m_{\phi}\approx\, 1\times10^{-17}- 8.3\times10^{-13}\,$eV}. Part of this range has thus far remained 
comparatively unexplored 
for scalar UBDM, since it is out of reach for both state-of-the-art atomic clock searches (e.g.\,\cite{KennedyPRL2020,BeloyNature2021})
and a search with a 
gravitational-wave detector \cite{Vermeulennature2021}. In addition to scalar UBDM, we search for pseudoscalar UBDM in the range \mbox{$m_{\phi}\approx\,5\times10^{-18}- 4.1\times10^{-13}\,$eV} employing the sensitivity of  Experiment 1 to the QCD axion and improving over the results of previous laboratory searches in  a part of this mass range. 


\emph{UBDM detection approach ---}
In the presence of scalar UBDM-SM interactions which are first order in the UBDM field \footnote{See \cite{HeesPRD2018,Banerjee:2022sqg} for phenomenology of second-order couplings.},  FCs such as $\alpha$, $m_{\rm e}$ and $\Lambda_{\rm {QCD}}$ may acquire time-dependent components: 
\begin{equation}
\label{eq:alphaVar}
\alpha(t)=\alpha_0\Big[1+d_{e}\frac{\phi(t)}{M_{\rm{Pl}}}\Big]\,,
\end{equation}
\begin{equation}
\label{eq:mVar}
m_{\rm e}(t)=m_{\rm{e,0}}\Big[1+d_{\rm{m_{e}}}\frac{\phi(t)}{M_{\rm{Pl}}}\Big]\,,
\end{equation}
\begin{equation}
\label{eq:nuclearVar}
\Lambda_{\rm{QCD}}(t)=\Lambda_{\rm{QCD,0}}\Big[1+d_{\rm g}\frac{\phi(t)}{M_{\rm{Pl}}}\Big]\,.
\end{equation}
\noindent Here $m_{\rm e,0}$,~$\alpha_{\rm 0}$ and $\Lambda_{\rm QCD,0}$ are the time-averaged values of the constants, $\phi(t)=\phi_{0}\sin{(2\pi f_{\rm C}t})$ is the UBDM field of amplitude $\phi_{0}=m_{\phi}^{-1}\sqrt{2\rho_{\rm DM}}$, where \mbox{$\rho_{\rm DM}\approx 3 \cdot 10^{-6}$}\,eV$^4$ is the estimated local galactic UBDM density \cite{KimballBook}, $M_{\rm{Pl}}=\sqrt{\frac{\hbar c}{8\pi G_{N}}}=2.4\times10^{18}$~GeV is the reduced Planck mass (with $G_{\rm N}$ being the Newtonian gravitational constant), 
and ~$d_{\rm e}$,~$d_{\rm m_e}$,and $d_{\rm g}$  are the respective couplings.

If instead, UBDM is due to the  pseudoscalar QCD-axion field,  because of axion-pion mixing, the oscillating axion background is expected to induce a temporal dependence of the pion mass~\cite{Ubaldi:2008nf}, and thus add an oscillating component to the nucleon masses and the nuclear $g$-factor~\cite{Kim:2022ype}. 
In this case, 
the proton  mass $m_{\rm p}$, and the nuclear $g$-factor for $^{87}$Rb ($g_{\rm nuc}$) can be written as \cite{Kim:2022ype,Flambaum:2006ip}:

\begin{equation}
m_{\rm p} (t)= m_{\rm{p,0}}\Big[1- \frac{6.6\times 10^{-3}}{f_{\phi}^2} \phi(t)^2\Big]\,,
\label{eq:mp_var}
\end{equation}
\begin{equation}
g_{\rm nuc}(t) = g_{\rm nuc,0}\Big[1 + \frac{2.6\times 10^{-3}}{f_{\phi}^2} \phi(t)^2\Big]\,,
\label{eq:gnuc_var}
\end{equation}

\noindent where, $m_{\rm{p,0}}$, $g_{\rm nuc,0}$ are the time-averaged values of the parameters, 
and $1/f_{\phi}$ is the QCD-axion coupling with the SM gluon fields, with $f_{\phi}$ being the QCD-axion decay constant.

The essence of our UBDM detection approach is to compare the frequencies of two systems (atomic vs. acoustic resonance) that depend on oscillating parameters differently. Generally, a change of a constant $\lambda$ by $\delta \lambda$ may change the 
resonance frequency, 
$f_{i}$, by $\delta f_{i}$, which can be quantified with a sensitivity coefficient $K_{{i}}^{\rm{\lambda}}=(\delta f_{i}/f_{i})/(\delta \lambda/\lambda_0)$ \cite{KozlovADP2019}. In frequency comparison of two systems $i$ and $j$, done for example by tuning one frequency close to the other, so that $f_{i} \approx f_{j}=f$, the difference $\delta f=\delta f_{i}-\delta f_{j}$  will also change with $\delta \lambda$ as long as the two oscillators exhibit different sensitivity to $\lambda$. The fractional change can be written as 
$\delta f/f=(K_{i}^{\rm{\lambda}}-K_{j}^{\rm{\lambda}})\delta \lambda/\lambda_{\rm 0}$, or assuming $n$ FCs 
changing:
%
%
%
\begin{equation}
\begin{aligned}
\label{eq:df}
\frac{\delta f}{f}=\sum_{n}\big(K^{\lambda_n}_{i}-K^{\lambda_n}_{j}\big)\frac{\delta\lambda_{n}}{\lambda_{\rm{0,n}}}.
\end{aligned}
\end{equation}
Equation\,\eqref{eq:df} is applied in comparing the frequency $f_{\rm Q}$ of a quartz-crystal oscillator to: i) the ground-state hyperfine resonance frequency $f_{\rm HF}$ in $^{87}$Rb  (comparison 1), and ii) the frequency $f_{\rm Dy}$ of an rf electronic transition  in $^{164}$Dy (comparison 2). The relevant sensitivity coefficients  are given in Table \ref{Table:Ktable}. 

The quartz frequency $f_{\rm Q}$ depends on 
$\alpha$, $m_{\rm e}$ and the nuclear mass 
$m_{\rm N}\propto A \cdot \Lambda_{\rm QCD}$\cite{CampbellPRL2020}, where $A$ is the mass number. An  atomic hyperfine frequency $f_{\rm HF}$ depends primarily on $\alpha$, $m_{\rm e}$ and  $m_{\rm p}\propto\Lambda_{\rm QCD}$ but with different sensitivities 
compared to $f_{\rm Q}$.
Thus, comparison 1 allows to probe oscillations of $\alpha$, $m_{\rm e}$ and $\Lambda_{\rm QCD}$ within the assumption of scalar couplings. (The frequency $f_{\rm {HF}}$ depends additionally on the quark masses with sensitivity coefficients $\ll 1$  \cite{FlambaumPRD2004}; these contributions are omitted here.)
Comparison 1 is one of few ways to probe oscillations of the nuclear mass~\cite{OswaldPRL2021,AntypasQST2021}, 
and it extends the investigated frequency range for FC oscillations
of a previous search based on  a quartz/H maser comparison \cite{CampbellPRL2020}. Applying Eq.\,\eqref{eq:df} with the use of Eqs.\,(\ref{eq:alphaVar}-- 
\ref{eq:nuclearVar}) and the values in Table \ref{Table:Ktable}, one obtains for the fractional frequency oscillations due to a scalar UBDM field: 
\begin{equation}
\begin{aligned}
\label{eq:dfRbQuartz}
\frac{\delta f_{\rm{HF}}-\delta f_{\rm{Q}}}{f}&=2.34\,\frac{\delta\alpha}{\alpha_0}+\frac{1}{2}\,\frac{\delta m_{e}}{m_{e,0}}-\frac{1}{2}\,\frac{\delta\Lambda_{\rm{QCD}}}{\Lambda_{\rm{QCD,0}}}\\
&= (2.34\,d_{e}+\frac{1}{2}\,d_{m_{e}}-\frac{1}{2}\,d_{g})\frac{\phi(t)}{M_{\rm{Pl}}}\,,
\end{aligned}
\end{equation}

\noindent We further consider comparison 1 via the quadratic coupling of the QCD-axion field, and make use of Eq.\,\eqref{eq:mp_var},\, \eqref{eq:gnuc_var} and Table \ref{Table:Ktable} to
write Eq.\,\eqref{eq:df} as 
\begin{equation}
\begin{aligned}
\label{eq:dfRbQuartzQCD}
\frac{\delta f_{\rm{HF}}-\delta f_{\rm{Q}}}{f}&=\frac{\delta g_{\rm nuc}}{g_{\rm nuc,0}} - \frac{1}{2}\frac{\delta m_{\rm p}}{m_{\rm{p,0}}} =\frac{ 5.9\times 10^{-3}}{f_\phi^2}\phi(t)^2.
\end{aligned}
\end{equation}
\noindent We see 
that, due to the quadratic coupling of $\phi(t)$, oscillations of $\delta f/f$ would appear at twice the UBDM particle's Compton frequency.

\begin{table}[h]
\caption{\small{Assumed 
fractional sensitivities of oscillator/transition frequencies to different FCs,
relevant for scalar and pseudoscalar interactions. 
}}
\begin{ruledtabular}
\begin{tabular}{c||c c c|c c|c}

&\multicolumn{2}{c}{\text{Scalar}}&&\multicolumn{2}{c|}{\text{Pseudoscalar}}\\

\hline\hline
&\textrm{$K^{\rm \alpha}$}&\textrm{$K^{\rm m_e}$}&\textrm{$K^{\rm \Lambda_{\rm QCD}}$}&\textrm{$K^{\rm{m_p}}$}&\textrm{$K^{\rm{g_{nuc}}}$}&\textrm{Ref.}\\
\colrule
Quartz & 2 & 3/2 & $-1/2$ & $-1/2$ &-&\cite{CampbellPRL2020} \\[0.2cm]
$^{87}$Rb & 4.34 & 2 & $-1$ & $-1$ & 1&\cite{FlambaumPRD2004} \\[0.2cm]
$^{164}$Dy & $2.6\times10^6$ & 1 & - & - & - & \cite{DzubaPRA2003}
\end{tabular}
\end{ruledtabular}
\label{Table:Ktable}
\end{table}


%

If no oscillations of $\delta f/f$ are detected,  Eq.\,\eqref{eq:dfRbQuartz} can be used to constrain the couplings $d_{\rm e}$, $d_{\rm m_e}$, $d_{\rm g}$ 
for scalar UBDM, and Eq.\,\eqref{eq:dfRbQuartzQCD} the coupling $1/f_{\phi}$ for QCD-axion UBDM. Note that, although $1/f_{\phi}$ and $m_{\phi}$ are related for the QCD axion~\cite{Peccei:1977ur,Peccei:1977hh,Weinberg:1977ma,Wilczek:1977pj,Kim:1979if,Shifman:1979if,Zhitnitsky:1980tq,Dine:1981rt}, here we treat them as independent quantities 
and estimate the reach of the experiment
within a more general class of models.


In comparison 2, we benefit from using an electronic transition in Dy exhibiting extreme sensitivity to changes of $\alpha$.
The transition is between two nearly degenerate, excited energy levels: the $\rm{4f^95d^26s}$ and $\rm{4f^{10}5d6s}$ levels. Primarily due to the small transition  frequency $f_{\rm Dy}$ (754 MHz in $^{164}$Dy) and significant relativistic effects, the transition has large fractional sensitivity to $\alpha$ changes~\cite{DzubaPRA1999, DzubaPRA2003, DzubaPRA2008}. 
It has been employed for searches of linear-in-time drift of $\alpha$ and  
$\alpha$ oscillations \cite{VanTilburgPRL2015} on time scales from several seconds to years. Here we extend the search in the previous work \cite{VanTilburgPRL2015}, primarily addressing a frequency range for the FC oscillations (100\,mHz-200\,Hz) that was not explored in \cite{VanTilburgPRL2015}. We focus on scalar UBDM and oscillations of $\alpha$, and write, analogously to Eq.\,\eqref{eq:dfRbQuartz}:
\begin{equation}
\label{eq:dfDyQuartz}
\frac{\delta f_{\rm{Dy}}-\delta f_{\rm{Q}}}{f} 
\approx2.6\times 10^6\,\frac{\delta\alpha}{\alpha_0}\\
 =2\times10^6\,d_e\frac{\phi(t)}{M_{\rm{Pl}}}\,.
\end{equation}
\noindent Since Dy sensitivity to $\alpha$ 
is vastly larger than that of quartz, the comparison 2 has practically no dependence on the quartz frequency oscillating with $\alpha$.  

\emph{Apparatus.---} In both experiments, spectroscopy of the respective rf transitions is implemented, probing atoms with an rf field produced from a quartz oscillator. The apparatus are described in detail in the Sup. Mat.

\begin{figure}[t]
    \centering
    \includegraphics[width=\columnwidth]{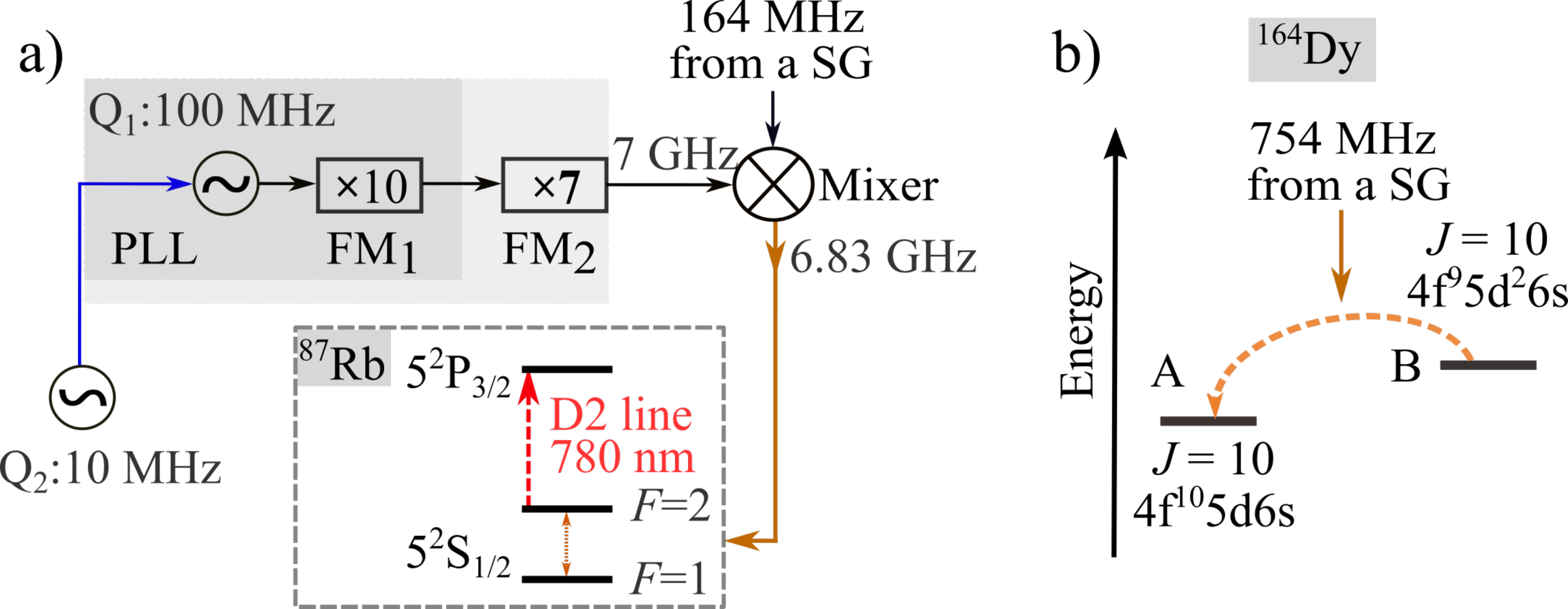}
 \caption{ 
 \small{ a) Schematic of the setup to produce the  rf signal probing the $^{87}$Rb hyperfine transition at 6.83\,GHz. b) The electric-dipole
 rf transition between the excited $^{164}$Dy levels `B' and `A', both having the same angular momentum $J$. Abbreviations: PLL: phase-locked loop; FM: frequency multipliers; SG: signal generator. 
 }
}
    \label{fig:Schematics}
\end{figure}

Apparatus I implements vapor-cell-based spectroscopy of the $^{87}$Rb hyperfine transition, employing the optical-microwave double-resonance technique \cite{Demtroder}. A Rb transition is excited with a microwave field produced by mixing the output of a 100-MHz, oven-controlled, stress-compensated (SC)-cut quartz oscillator (Q$_{\rm 1}$ in Fig.\,\ref{fig:Schematics}), multiplied to 7\,GHz, with the signal from a function generator. The resulting frequency is close to the hyperfine-resonance frequency of $\approx6.83$\,GHz. To 
reduce low-frequency noise, the  oscillator Q$_{\rm 1}$ is phase-locked to another oscillator (Q$_{\rm 2}$ in Fig.\,\ref{fig:Schematics}), that exhibits higher long-term frequency stability compared to Q$_{\rm 1}$. The oscillator Q$_{\rm 2}$ is also oven-controlled and constructed around a SC-cut, quartz crystal. The phase-locked loop has a measured bandwidth of $\approx$3\,Hz. Thus, for $f_{\rm{ C}}\leq3$\,Hz the dependence on the oscillating parameters ($\alpha$,\,$m_{\rm{e}}$\,and $m_{\rm N})$ is determined by Q$_{\rm 2}$, and for $f_{\rm{C}}> 3$~Hz by Q$_{\rm 1}$. The same dependence on the FCs is assumed for both. 


Experiment 2 utilizes an atomic beam setup for spectroscopy of the Dy rf transition (see \cite{Leefer2013_PRL} and references therein). 
The atoms are prepared in the metastable state $\rm{4f^95d^26s}$ (labeled `B' in Fig.\,\ref{fig:Schematics}) via a two-step laser excitation and subsequent decay. 
The signal from a signal generator at $\approx 754$\,MHz is used to produce an electric field that induces transitions to the $\rm{4f^{10}5d6s}$ state (state `A'), whose subsequent decay is monitored via fluorescence as a means to observe the B$\rightarrow$A transition, with an observed linewidth of $\approx$ 50\,kHz. The time base for the generator is provided by an internal oven-controlled, SC-cut, quartz oscillator. 

Frequency-modulation spectroscopy \cite{Demtroder} is implemented in both experiments, to improve detection sensitivity of the atomic excitations. For this, the respective rf drive is modulated in frequency and phase-sensitive detection of the spectroscopy signal is done.
%
%


\emph{Data acquisition and analysis---} In the two experiments, the spectroscopy signal was repetitively acquired for several values of the rf-modulation    frequency. In apparatus I, we found that the experimental parameters providing optimal sensitivity are different for different ranges of the frequency $f_{\rm{C}}$ (see Sup. Mat.). Thus, in the
low-frequency run, we recorded many 4-h-long time series for a total of 600\,h, with a sampling rate of 41.7\,Sa/s; while in the high-frequency run, we acquired a sequence of 10-min-long time series, for a total of 144\,h, sampling the signal at $406.5$\,Sa/s. Data taking for Experiment 2 was a total of 12\,h, with the spectroscopy signal sampled at 406.5\,Sa/s and recorded in three successive, 4-h-long time series. 

From the recorded time series, power spectra were computed and averaged. The corresponding amplitude spectra 
were investigated for possible signatures of oscillations that would appear as amplitudes 
in frequency bins of the spectra, that are 
greater than a threshold for detection. This threshold
is determined by the random noise in the vicinity of the bins, and set to a 95\% confidence level, accounting for the look-elsewhere effect \cite{Scargle1982} (see Sup. Mat.). We checked this set threshold by injecting artificial signals to the recorded time series, and looking at the size of the respective amplitudes in the computed spectra
(see Sup. Mat.). 

\begin{figure}[t]
    \centering
    \includegraphics[width=\columnwidth]{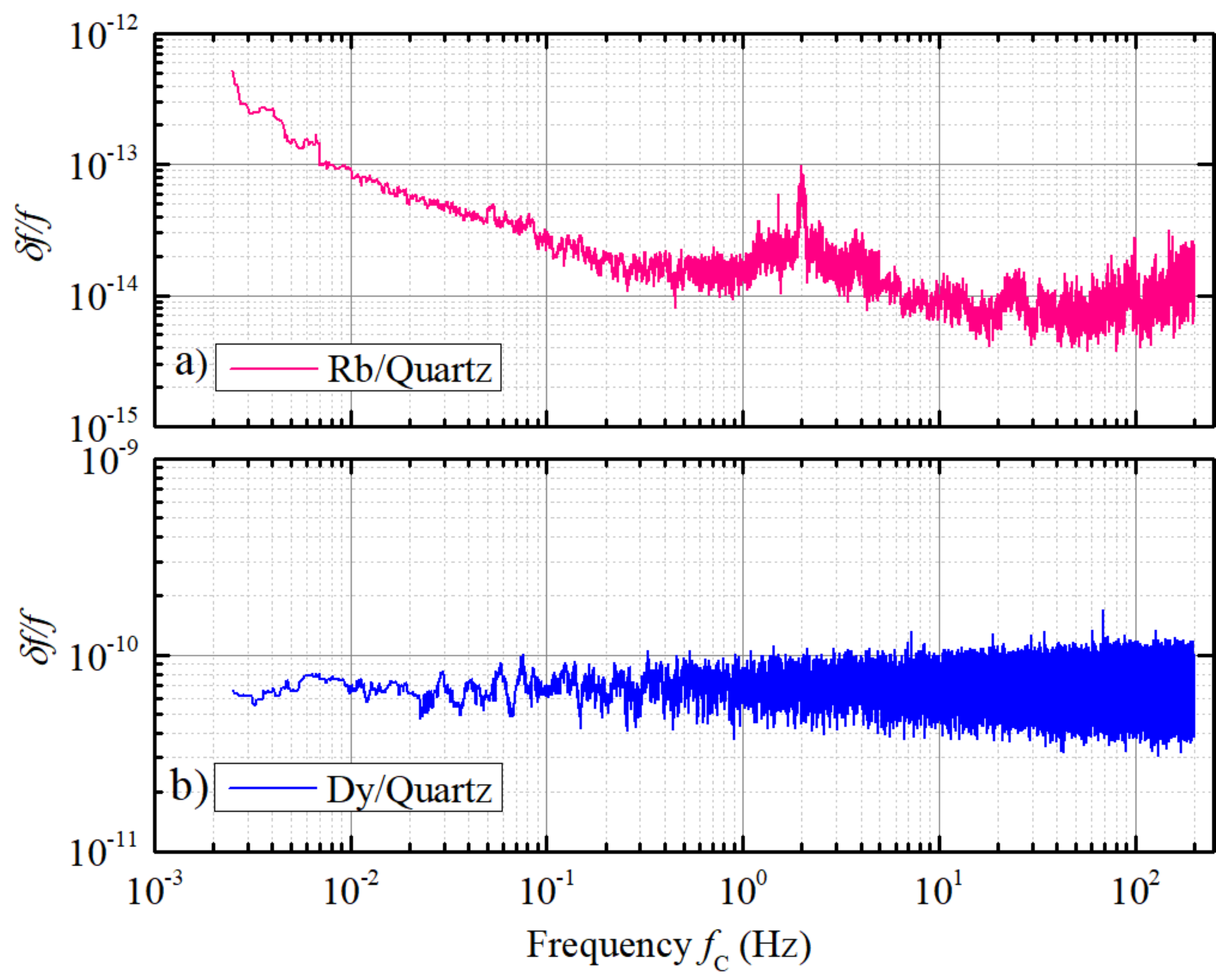}
 \caption{\small{Exclusion limits on fractional frequency oscillations $\delta f/f$ at 95$\%$ C.L. a) Experiment 1. The spectrum is produced by merging the spectra from the low- and high-frequency runs  at 5 Hz, i.e. the frequency where the respective FC detection sensitivities become equal. We do not provide limits in the frequency windows $50\pm0.25$\,Hz and $100\pm0.25$\,Hz. b) Experiment 2. The spectrum  exhibits no pronounced dependence on frequency, as the dominant noise source is shot noise in the detection of Dy transitions in the atomic beam.
}
}
    \label{fig:dff}
\end{figure}

 A total of ten peaks were observed to exceed the threshold for detection in the low-frequency run of Experiment 1 and 231 peaks were seen in the high-frequency run. In Experiment 2, 983 peaks were observed. All these spurious signals were checked 
 via: i)\,intercomparison of the averaged amplitude spectra acquired with different modulation frequencies; ii)\, cross-checks between the spectra acquired for the low- and high-frequency range runs (relevant in Experiment 1); iii)\, comparison of primary data sets against sets from auxiliary runs with an alternative signal generator (see Fig.\,\ref{fig:Schematics}). As an actual UBDM signal should persist in all these tests, eventually all spurious peaks were excluded from being UBDM candidates, allowing us to constrain the  spectra of the fractional frequency oscillations $\delta f/f $, as shown in Fig.\,(\ref{fig:dff}) \footnote{It is challenging to reliably compute a threshold for detection of oscillation in the $\delta f/f$ spectra for $f_{\rm C}<1$\,mHz (see Sup. Mat.). Here we provide constraints for $f_{\rm C}\geq 2.5$\,mHz  }.


\begin{figure}[t]
    \centering
    \includegraphics[width=\columnwidth]{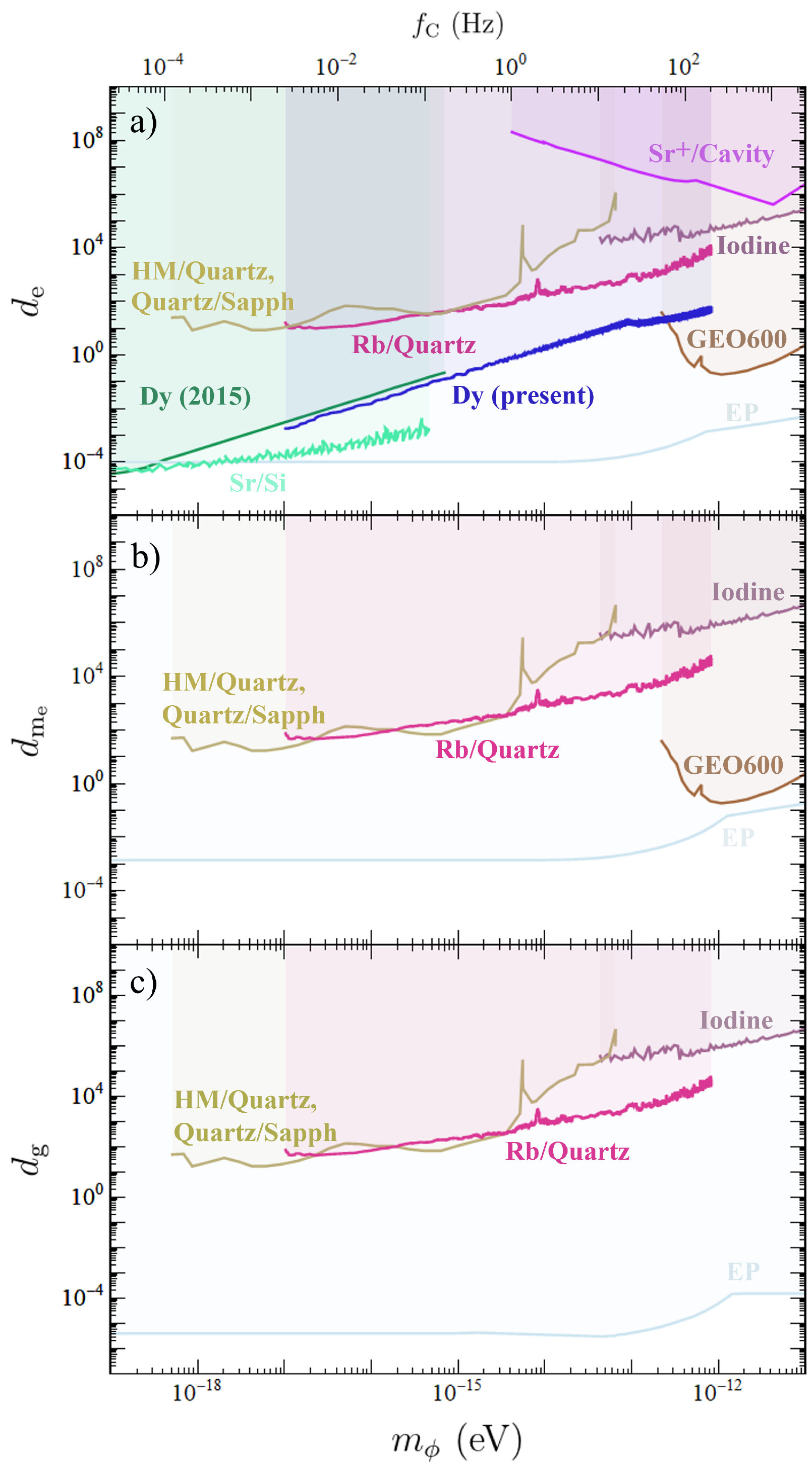}
 \caption{\small {
 Constraints on the UBDM couplings to $\alpha$, $m_{\rm{e}}$, and $\Lambda_{\rm {QCD}}$  from the present work (Rb/Quartz and Dy/Quartz), shown  at the $95\%$ C.L., alongside constraints from other experiments. 
Dy (2015):\cite{VanTilburgPRL2015},
Sr/Si: \cite{KennedyPRL2020}, Sr$^+$/cavity: \cite{AharonyPRD2021}, Iodine:\cite{OswaldPRL2021},
GEO600: \cite{Vermeulennature2021},
Hydrogen Maser (HM)/Quartz-Quartz/Sapphire:
\cite{CampbellPRL2020},
EP:\cite{Touboul:2017grn,Wagner:2012ui,Smith:1999cr}. The limits from \cite{CampbellPRL2020} are plotted considering the respective parameters independently, and multiplying by
a factor $\times 4.4$ to account for stochasticity of UBDM in that work, as it was done for the Rb/Quartz and Dy/Quartz data (see text). 
}}
    \label{fig:de}
\end{figure}

\emph{Constraints on UBDM couplings---}We use the constraints on $\delta f /f $ (Fig.~\ref{fig:dff}) and Eq.~(\ref{eq:df}),
(\ref{eq:dfRbQuartz}),(\ref{eq:dfDyQuartz}) to bound the UBDM couplings to $\alpha$, $m_{\rm e}$ and $\Lambda_{\rm QCD}$ (Fig.~\ref{fig:de}). To do this, we assume that the respective coupling dominates the UBDM-SM interaction. In addition, we consider the stochastic nature of the UBDM field \cite{Centers2021} and apply  a correction to the bounds to account for reduction in UBDM detection sensitivity that becomes appreciable at oscillation frequencies $f_{\rm{C}} <Q/T$ \footnote{To within a factor of $2\pi$ \cite{GramolinPRD2022}}, where $Q\approx 1.1\times10^6$  is the Q-factor of the UBDM field within the standard galactic UBDM halo scenario \cite{GramolinPRD2022}, and $T$ is the total measurement time; $T=864$~h and $12$~ h for experiments I and II, respectively. Applying the analysis method of \cite{PelssersThesis} we find that the bounds from experiments I and II become weaker by a factor of $\approx\, 4.4$ below $\approx\, 1$\,Hz and $\approx$\,50\,Hz, respectively \footnote{This $\approx\, 4.4$ correction factor may be conservative, compared to the factor $\approx\, 3$ of \cite{Centers2021}}.


The bounds on $d_{\rm e}$, $d_{\rm m_e}$ and $d_{\rm g}$ from the Rb/quartz comparison improve on previous results by as many as $\times 100$ times in the range 1$-$200\,Hz. Within the whole range investigated (2.5\,mHz$-$200\,Hz), a variety of experiments directly probe for oscillations of $\alpha$ and $m_{\rm e}$, as seen in Fig.\,\ref{fig:de}a and Fig.\,\ref{fig:de}b. Few experiments however, probe a hyperfine resonance (as we do in this work), and are sensitive to oscillations of the strong force (Fig.\,\ref{fig:de}c).

\begin{figure}[t]
    \centering
    \includegraphics[width=\columnwidth]{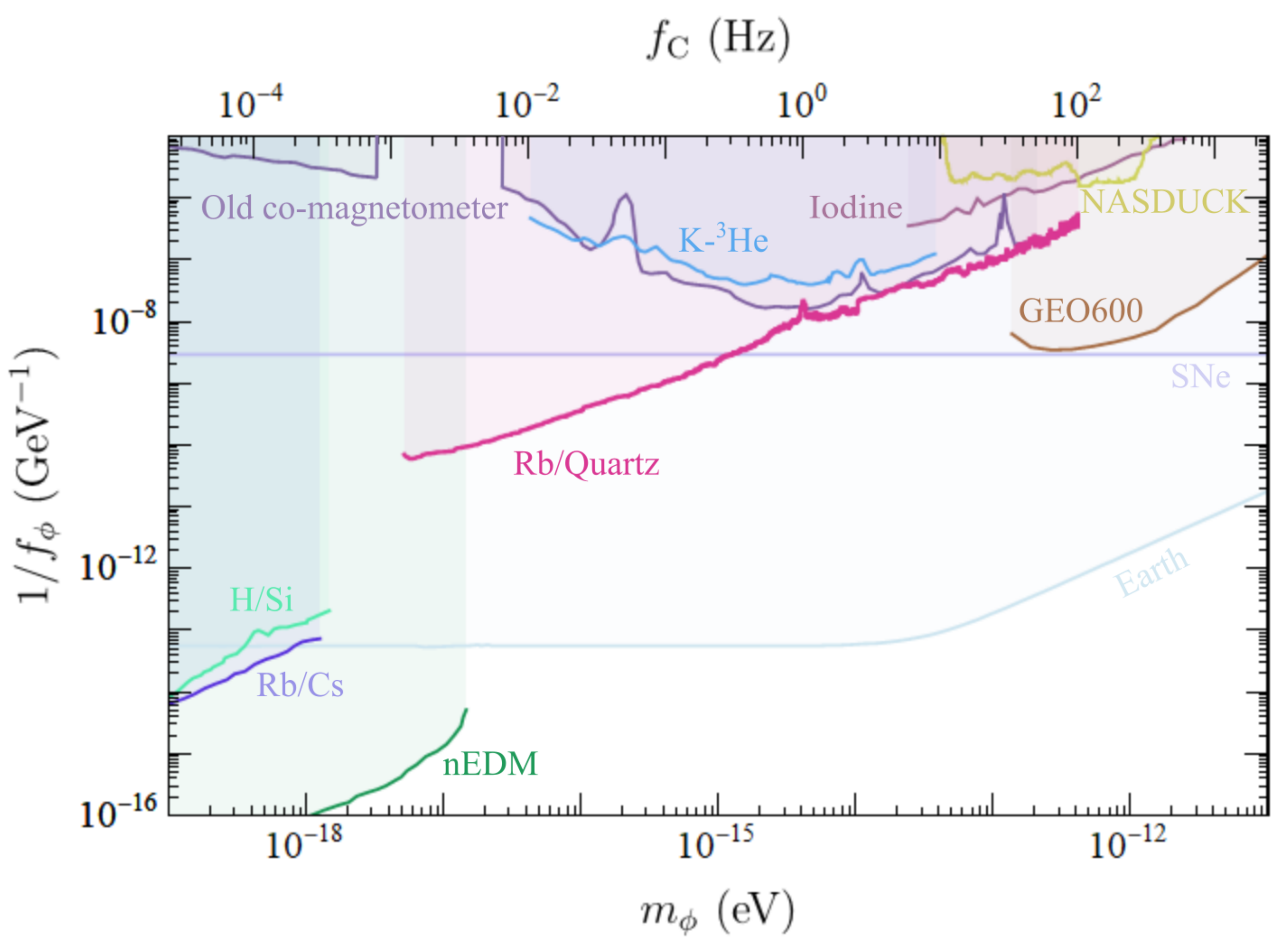}
 \caption{\small{ Constraints on the QCD axion-gluon coupling. Rb/Cs \cite{HeesPRL2016}, H/Si \cite{KennedyPRL2020}, nEDM \cite{Abel:2017rtm}, Iodine \cite{OswaldPRL2021}, GEO 600 \cite{Vermeulennature2021}, SNe \cite{Raffelt:2006cw}, Co-magnetometer\cite{Bloch:2019lcy}, NASDUCK \cite{Bloch:2021vnn}, K-$^3$He \cite{LeeArxiv2022}, Earth \cite{Hook:2017psm}.
 }}
\label{fig:1overf}
\end{figure}

A more stringent bound on  $d_{\rm e}$ is provided by the Dy 
experiment. 
The limit $\delta f/f \approx 8\times 10^{-11}$ translates to a limit
$\delta \alpha/\alpha\approx 3\times10^{-17}$
and a bound on $d_{\rm e}$ that improves on previous results by as many as three orders of magnitude. Having UBDM detection capability up to the frequency of the observed transition linewidth ($\approx 50 $~kHz) the Experiment 2 is used to explore a region  between the upper-frequency end in  state-of-the-art atomic clock searches (e.g.\, \cite{KennedyPRL2020}) and the low-frequency end of the GEO600 search \cite{Vermeulennature2021}.  

The assumption of pseudoscalar UBDM allows one to interpret the $\delta f/f$ limits from Experiment 1 as limits on the QCD axion-gluon 
coupling $1/f_{\rm{\phi}}$ (Fig.\,\ref{fig:1overf}) via Eq.\,\eqref{eq:dfRbQuartzQCD}.  Here as well a correction is made to account for the stochasticity of UBDM; it amounts to a calculated degradation of the limit of Fig.\,(\ref{fig:dff}) by a factor of $\approx 2.5$ in the sub-Hz region. 
Our constraints on $1/f_{\rm{\phi}}$ (Fig.\,\ref{fig:dff}) improve on those from tabletop experiments probing the effects of an axion coupling via atomic magnetometry \cite{ Bloch:2019lcy,Bloch:2021vnn}. They also surpass astrophysical limits \cite{Raffelt:2006cw} in the frequency range below 200 mHz. 

\emph{Conclusions and outlook ---} Our bounds on scalar and pseudoscalar UBDM interactions represent significant improvement over previous work in part of the explored mass range. While the limits on scalar couplings to the $\alpha$, $m_{\rm e}$ and $d_{\rm g}$ from EP-violation searches are more stringent, direct searches for oscillations in these constants offer important cross-checks. 
In addition, as discussed in~\cite{Oswald:2021vtc,Banerjee:2022sqg}, 
if the scalar UBDM has some non generic coupling to the SM, then bounds from the EP-violation/fifth-force experiments may be suppressed by a factor $\mathcal{O}(10^{3})$ and may become comparable to that of FC-oscillation searches~\cite{Wrestle2}. 



The recent work \cite{Kim:2022ype} pointing to oscillatory effects in nuclear parameters in the presence of the QCD axion, extends the physics reach of 
apparatus used thus far to check for FC oscillations. As we show here,
this opens a way to probe pseudoscalar UBDM with sensitivity that is, in a certain mass range, far greater than that in setups designed to search for previously considered pseudoscalar-field
observables. 

This possibility motivates further apparatus improvements, for example, in probing the hyperfine resonance. The present Rb/quartz frequency comparison is at the $10^{-12}/\sqrt{\tau}$ level in the short-term 
(measurement time $\tau< 10$\,s); this is $\approx$10 times lower than that reported for a vapor-cell-based Rb clock \cite{BandiIEE2014}.
Long-term stability can be improved by optimization of the parameters of the Rb-vapor-cell setup \cite{BandiIEE2014}. 
Because the stability of a quartz oscillator degrades at such long time scales, it would be necessary to replace it with  a microwave signal derived from an optical atomic clock or an optical cavity \cite{KennedyPRL2020}. 
Together with a long data-taking campaign, such improvements could extend the reach of an experiment by orders of magnitude, and probe for the QCD-axion further beyond the level allowed by atomic magnetometry and astrophysical observations.\\


We thank W.\,Ji for discussions and N.\,L.\,Figueroa, D.\,Kanta, U.\,Rosowski and M.\,Hansen for help with the project. This work was supported by the European Research Council (ERC) under the European Union
Horizon 2020 research and innovation program (project YbFUN, grant agreement No 947696) and by the DFG Project ID 390831469: EXC 2118 (PRISMA+ Cluster of Excellence). The work of AB is supported by the Azrieli foundation.


\clearpage
\bibliographystyle{apsrev4-1}
\bibliography{bibliography.bib}

\newpage

\newpage

\section*{Supplemental Material}

\subsection{Apparatus}
\label{"subsec:SI-apparatus"}
\begin{figure}[t]
    \centering
    \includegraphics[width=\columnwidth]{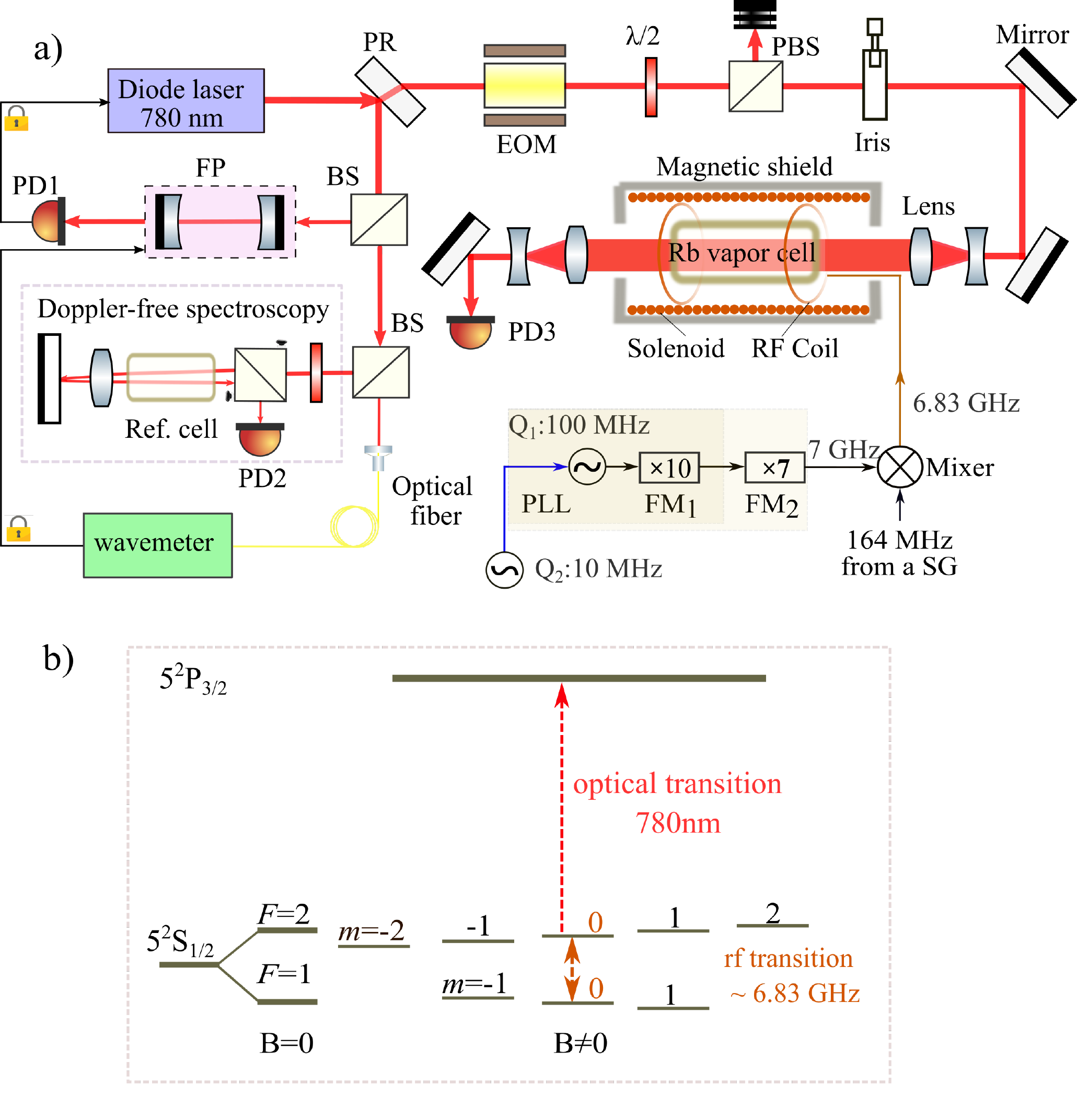}
 \caption{ a) 
 Experimental setup for Experiment 1. PR: partial reflector; EOM: electro-optic modulator; $\lambda/2$: half-wave plate; (P)BS: (polarizing) beam splitter; FP: Fabry-P\'erot optical cavity; PD: photodetector. PLL: phase-locked loop; FM: frequency multipliers. Q1, Q2: quartz oscillators. b) Optical-microwave double resonance for ${}^{87}$Rb. The red dashed line indicates optical pumping and the orange dashed line shows hyperfine transition between the clock state.
 }
 
    \label{fig:Rb_setup}
\end{figure}

\emph{Experiment 1 ---} A schematic of the Rb/quartz setup is shown in Fig.\,\ref{fig:Rb_setup}\,a). The optical-microwave double resonance technique is applied to Rb vapor to look for the effects of ultralight bosonic dark matter (UBDM). A cylindrical cell (25 mm long, with a 25 mm diameter) 
containing natural-abundance Rb-metal vapor and buffer gasses (14\,mbar argon and 10\,mbar nitrogen) is placed in a cylindrical plastic tube that holds two wire loops which provide rf magnetic field 
to the atoms. This assembly lays inside a solenoid that provides a static magnetic field 
in the range 0.5-1\,$\rm{\mu}$T, to resolve the Zeeman sublevels of the ${}^{87}$Rb ground hyperfine levels. A pair of heater tapes are wrapped around the solenoid to heat the Rb vapor in the range 55-65 $^{\circ}$C. The whole assembly is placed inside a single-layer magnetic shield. 

Light from a diode laser is used to drive the ${}^{87}$Rb D2 transition at 780\,nm. 
In order to enhance the sensitivity of the apparatus, the laser frequency noise is actively reduced by stabilizing the laser frequency to a resonance of a Fabry-P\'erot cavity. The cavity is in turn stabilized to the reading of a wavemeter (High Finesse WS8-2), so that the the long-term drift of the laser frequency is suppressed and the laser frequency remains stabilized to a point where the signal from the hyperfine resonance is optimal. Another setup, allowing Doppler-free spectroscopy 
of the D2 line is used as an auxiliary frequency reference. 

The 780\,nm beam entering the vapor cell has a power of $\approx\,0.2$\,mW and $\approx$15 mm diameter. The power of the beam is stabilized with an electro-optic amplitude modulator and a proportional-integral-derivative controller (not shown in the schematic). 

The primary quartz oscillator in the setup (Q$_1$ in Fig.\,\ref{fig:Rb_setup}a) is a 100-MHz unit based on an oven-controlled, SC-cut crystal.  The oscillator is integrated in a device (NEL Frequency Controls O-CEGM-017DWEP-R-1\,GHz) that incorporates an analog multiplier to produce a 1-GHz output. The device is housed in a multiplier unit (NEL Frequency Controls N-DCN-SS702-000IR-7.00\,GHz) that brings the signal to 7\,GHz with use of analog multipliers. As mentioned in the main text, to improve low-frequency noise performance, the 100-MHz oscillator is phase-locked with a bandwidth of $\approx$3\,Hz to another, more stable, 10-MHz oscillator (Q$_2$ in Fig.\,\ref{fig:Rb_setup}a), constructed around an oven-controlled, SC-cut quartz (NEL Frequency Controls O-CE1-0S19HR-N-E-N-R\,10.000\,MHZ).
The 7-GHz signal is mixed with use of a frequency mixer (Mini-Circuits ZX06-U742MH-S+) with a $\approx$164-MHz signal from a signal generator (SRS SG386) to produce a field at 6.83\,GHz that drives the $^{87}$Rb hyperfine transition. The generator's internal time base is another SC-cut quartz oscillator. The 164-MHz signal of the generator is frequency-modulated, as further explained below.



\begin{figure}[t]
    \centering
    \includegraphics[width=\columnwidth]{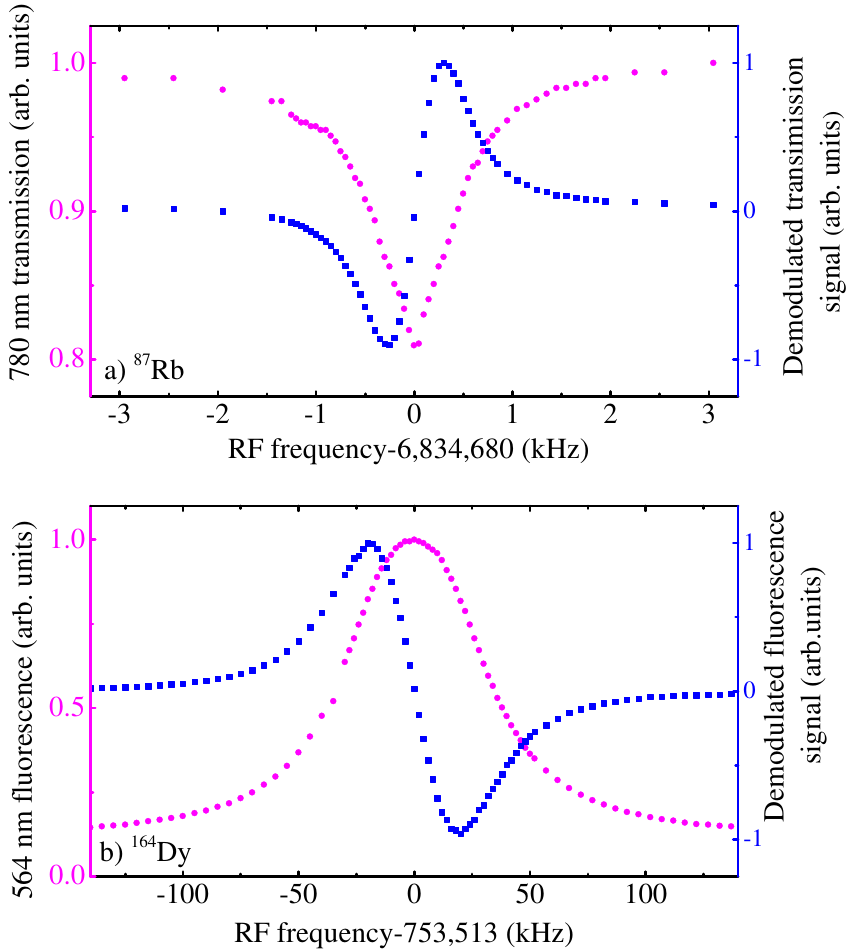}
    \caption{ a) Hyperfine resonance spectrum for ${}^{87}$Rb in the high-frequency run. The blue line shows the rf resonance, and the magenta line indicates the derivative lineshape acquired with the lock-in amplifier. b) rf resonance spectrum for ${}^{164}$Dy. The colors correspond to signals as in a).
    } 
    \label{fig:Magneticresonancespectra}
\end{figure}

The principle of the double-resonance technique is illustrated in Fig.\,\ref{fig:Rb_setup}\,b\,\cite{Bandi2012_APS}. Our  780-nm laser is tuned in frequency to excite the D2 transition from the $\rm{5^{2}S_{1/2}}$\,$\ket{F=2}$ ground level. 
Atoms are thus pumped to the other ground level, the  $\rm{5^{2}S_{1/2}}$\,$\ket{F=1}$ level, resulting in reduction of population of the $\ket{F=2}$ level. An rf field  at 6.83\,GHz drives the hyperfine $\ket{F=1,m=0}\xrightarrow{}\ket{F=2,m=0}$ clock transition between the two ground states, leading to an increase of the population of the $\ket{F=2,m=0}$ state, and resulting in decreased transmission of the 780\,nm light through the atomic sample. This transmission signal is a probe of the hyperfine resonance. The resulting spectrum, produced as the frequency of the rf field is swept around the hyperfine resonance,  is shown in Fig.\,\ref{fig:Magneticresonancespectra}\,a). Modulation of the rf-field frequency and demodulation with a lock-in amplifier provides a dispersion-shaped resonance signal, that is used as a frequency discriminator in the search for oscillating UBDM effects. The slope of the linear part of the dispersion-shape resonance is measured and used to calibrate the 
response, i.e. to convert the computed amplitude spectra described in the main text to $\delta f/f$ spectra.

\begin{figure}[t]
    \centering
    \includegraphics[width=\columnwidth]{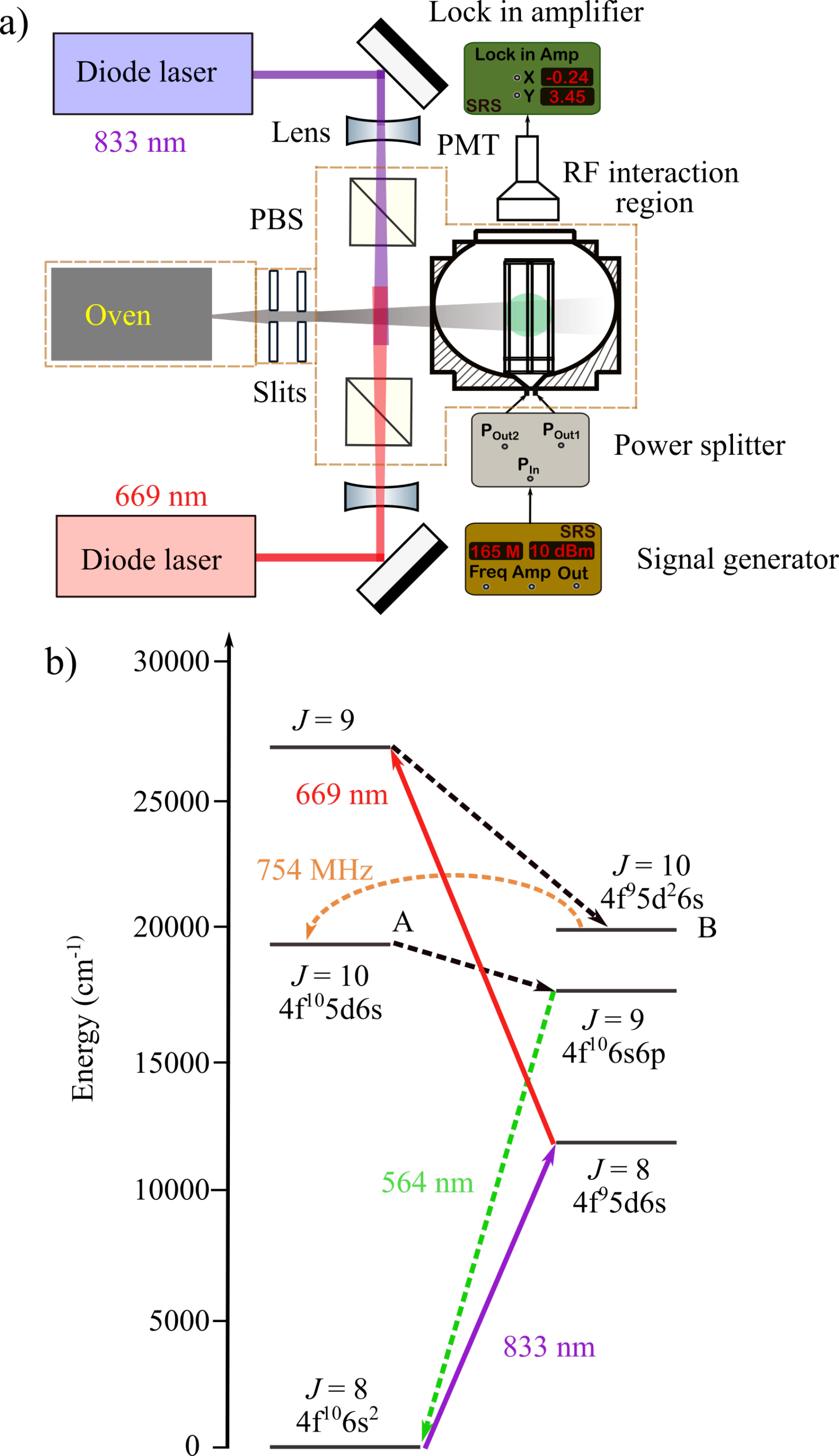}
    \caption{
    a) Apparatus schematic of Experiment 2. Components within orange dashed boundaries are in vacuum. b) 
    Partial energy level diagram of Dy. 
    The employed rf electric-dipole transition is between the nearly degenerate states, labeled `A' and `B'. Atoms are prepared in state B via two-step laser excitation followed by a spontanepous decay. Cascade decay from state A to the ground state yields fluorescence at 564\,nm (green dashed line).
    PMT: photomultiplier; PBS: polarizing beam splitter.
} 
    \label{fig:Dy_setup}
\end{figure}

\emph{Experiment 2 ---} A schematic of the Dy spectroscopy setup is shown in Fig.\,\ref{fig:Dy_setup}\,a. A thermal beam of Dy atoms, effusing from an oven heated to 1400\,K, is collimated with a pair of slits, and optically pumped via two-step excitation using light at 833\,nm and 669\,nm produced with diode lasers. This excitation is done with diverging laser beams in order to excite all atoms in the atomic beam \cite{NguyenPRA2000}. The excitation, shown in Fig.\,\ref{fig:Dy_setup}\,b, populates the state 
$\rm{4f^{10}5d6s}$\,$\ket{J=9}$ that spontaneously decays with a $30\%$ branching ratio into state B\,\cite{NguyenPRA2000}.
An ac electric field
at 754\,MHz induces an electric-dipole
transition between states B and A. The ac field is applied in the so-called interaction region, with parallel grids of 50-$
\mu$m-thick Be-Cu
wires constituting
electric field ``plates''. (A detailed description of the rf interaction region is given in \,\cite{Armanthesis}.) 
Atoms in state A (with lifetime of $\approx 8\,\mu$s\,\cite{BudkerPRA1994, Leefer2013_PRL})
exhibit cascade decay to the ground state $\rm{4f^{10}6s^{2}}$\,$\ket{J=8}$ emitting fluorescence at 564\,nm collected with a photomultiplier. This fluorescence is a probe of the rf resonance. 
The 754-MHz signal driving the Dy transition is produced with the same signal generator used in Experiment 1. As in Experiment 1, frequency modulation on the generator's output is done for phase-sensitive detection of the fluorescence signal Fig.\,\ref{fig:Magneticresonancespectra}b, using a lock-in amplifier. The demodulated signal, as in Experiment 1, serves as a frequency discriminator in the search for FC oscillations, whose linear part (as in Experiment 1) is used to calibrate the apparatus response.

\subsection{Data acquisition methods}
\label{"subsec:SI-Data acquisition methods"}

\emph{Experiment 1 ---} As mentioned in the main text, we find that experimental conditions providing optimal detection sensitivity are different for low and high frequencies. Therefore, we carry out separate low- and high-frequency data-taking runs. 

 The low-frequency run probes the range 2.5\,mHz-5\,Hz. The sensitivity is optimal for a vapor-cell temperature \mbox{$\approx$ 55$^{\circ}$C}, an optical intensity $\approx  1.12\,\rm{\mu}$W/mm$^{2}$, and an rf-drive power corresponding to  a hyperfine-resonance linewidth of $\approx$ 550\,Hz. 
 
A commercial 16-bit digitizer (PicoScope 5244D) is used for data acquisition. 
Data taking in the low-frequency run consists of recording the demodulated hyperfine-resonance signal in 4-h-long time series for a total of 600\,h, 
 at a sampling rate of $\approx 41.7$\,Sa/s. We alternate acquisition with the modulation frequency set one of two distinct values. This allows for intercomparison of spurious signals acquired in the respective spectra, and elimination of such signals as UBDM candidates. The 600-h recorded data involves 150 time series, of which 75 are obtained with modulation frequency of 177\,Hz and another 75 with 144\,Hz.

In the high-frequency run, we focus on the 5$-$200\,Hz frequency range. 
Here the rf-drive power yielding optimal sensitivity results in a  hyperfine-transition linewidth of $\approx$\,1 kHz; the Rb-cell temperature is $\approx$ 65$^{\circ}$C, and the optical intensity \mbox{$\approx$ $1.12\,\rm{\mu}$W/mm$^{2}$}.
The digitizer is set to ac-couple the signal (with cut-off at 1.5\,Hz) to eliminate noise at sub-Hz frequencies.
The data-taking run consists of recording the demodulated resonance signal in successive 10-min-long time series with 16-bit resolution, at a rate of $\approx 406.5$\,Sa/s. We alternate acquisition between three modulation frequencies: 873\,Hz, 884\,Hz and 895\,Hz. The total acquisition time is 144\,h, corresponding to 864 time series, evenly distributed among the three modulation frequencies. 

\emph{Experiment 2 ---} We acquire data in a run that covers the entire investigated $2.5$\,mHz$-$200\,Hz frequency range. We record a time series of demodulated rf-resonance signal in successive 4-h-long time series with 16-bit resolution at $\approx 406.5$\,Sa/s, for a total of 12\,h. Each 4-h-long time series is recorded with a dedicated modulation frequency for the rf field: 10,000\,Hz, 9,984\,Hz or 9,968\,Hz.

\subsection{Data analysis}
\label{"subsec:SI-Data analysis"}
\subsubsection{Obtaining  averaged power spectra}
\label{"subsubsec:SI-Data analysis-Averaged power spectra"}

\emph{Experiment 1 ---} The data acquired for the low- and high-frequency ranges of  Experiment 1 are analyzed separately. 

\begin{figure}[H]
    \centering
    \includegraphics[width=\columnwidth]{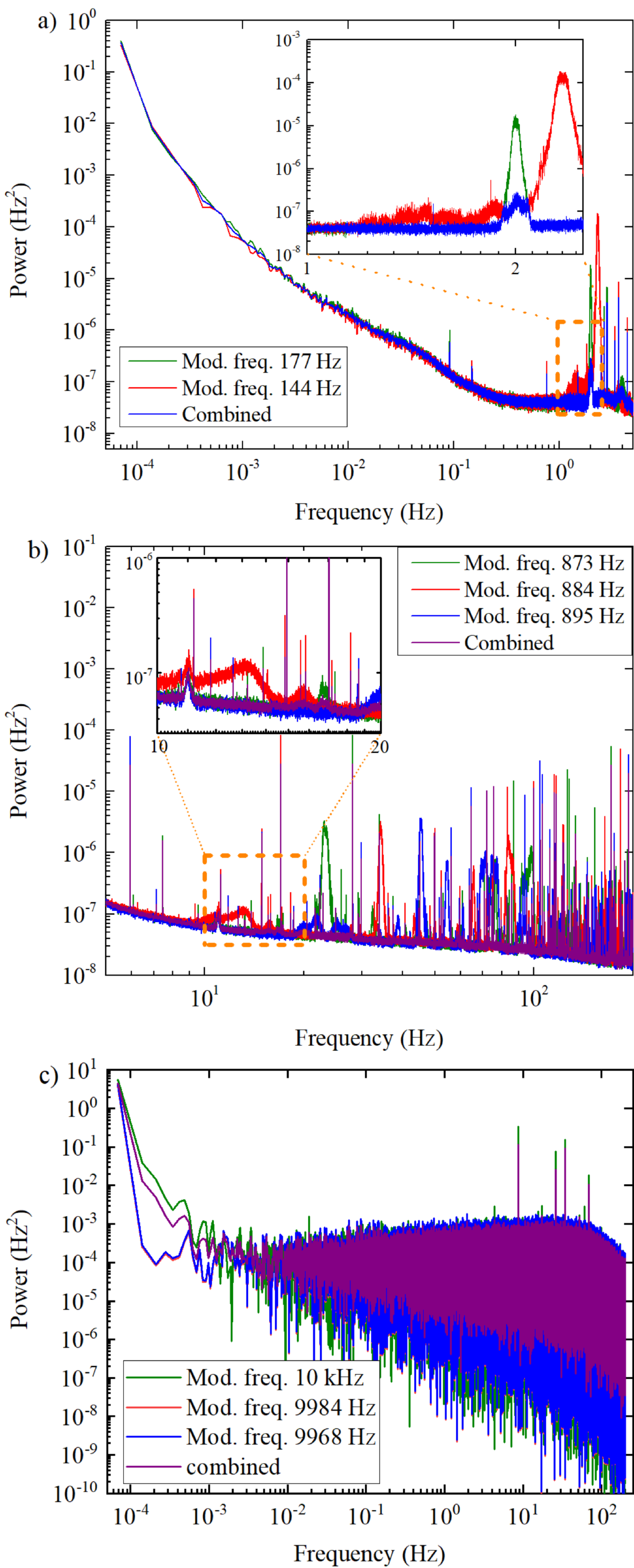}
    \caption{ Averaged power spectra computed from different measurements. a) Low-frequency spectra of Experiment 1. b) High-frequency spectra of Experiment 1.   c) Spectra of Experiment 2. See text for details. 
} 
   \label{fig:Powerspectra}
\end{figure}

For the low-frequency range, a Hann window is applied to the 4-h-long time series to avoid unwanted effects in the discrete Fourier Transform (DFT) which is subsequently performed on these data. From the DFT,  power spectra of the 4-h-long data are obtained. The power spectra acquired with different modulation frequencies are averaged separately \cite{AntypasPRL2019}. Therefore, two averaged power spectra are computed, corresponding to the two  modulation frequencies used in the low-frequency runs, as shown in Fig.\,\ref{fig:Powerspectra}\,a. 

Complications arise with further averaging of these two power spectra because 
excess noise power 
appears in some frequency ranges of the spectra, as shown in the inset of Fig.\,\ref{fig:Powerspectra}\,a. The origin of this excess power is predominantly pickup from laboratory sources, 
that drifts in frequency,
thus appearing as a broad-noise background in the average spectrum. Further combining the two averaged power spectra, would practically result in no improvement in the UBDM detection sensitivity within the region of such a broad excess-power background, as the noise in the resulting spectrum would be almost completely determined by the spectrum having the least noise. 

To carry out further analysis of the low-frequency data, we use the following protocol: 


\begin{enumerate}
    \item In frequency windows where no excess power is observed, all power spectra acquired with different modulation frequencies are directly averaged.
    \item In frequency windows where broad excess-power background is observed in one of the spectra (at least 40\% higher power compared to the vicinity of the window), only the data from the other(s) are retained and further averaged.
   	\item In the frequency windows where excess power is observed in all spectra, only the data from the spectrum with the least excess power are 
   	retained. 
\end{enumerate}

\noindent With application of this protocol, we arrive at a final averaged power spectrum
 that is indicated as `Combined' in Fig.\,\ref{fig:Powerspectra}\,a.



Analysis for the high-frequency range of Experiment 1 is done similarly. From the recorded 10-min-long time series, we compute averaged power spectra  (Fig.\,\ref{fig:Powerspectra}\,b), that we further combine using the above protocol to arrive at a final power spectrum that is labeled `Combined' in Fig.\,\ref{fig:Powerspectra}\,b.

\emph{Experiment 2 ---} The data analysis for Experiment 2 is more straightforward. This is because broad excess-noise backgrounds that complicate analysis in Experiment 1, are absent here. The three recorded time series (4-h-long each)  are computed to obtain respective power spectra, each belonging to one of the three modulation frequencies used in data acquisition. The three power spectra are further averaged together directly, yielding a final power spectrum, which is shown in Fig.\,\ref{fig:Powerspectra}\,c.

\subsubsection{Obtaining constraints on $\delta f/f$}
The procedure to obtain a 95$\%$-confidence-level (C.L.) threshold for detection of oscillations in the spectrum of $\delta f/f$  is the same for Experiment 1 and 2. 

\begin{figure}[t]
    \centering
    \includegraphics[width=\columnwidth]{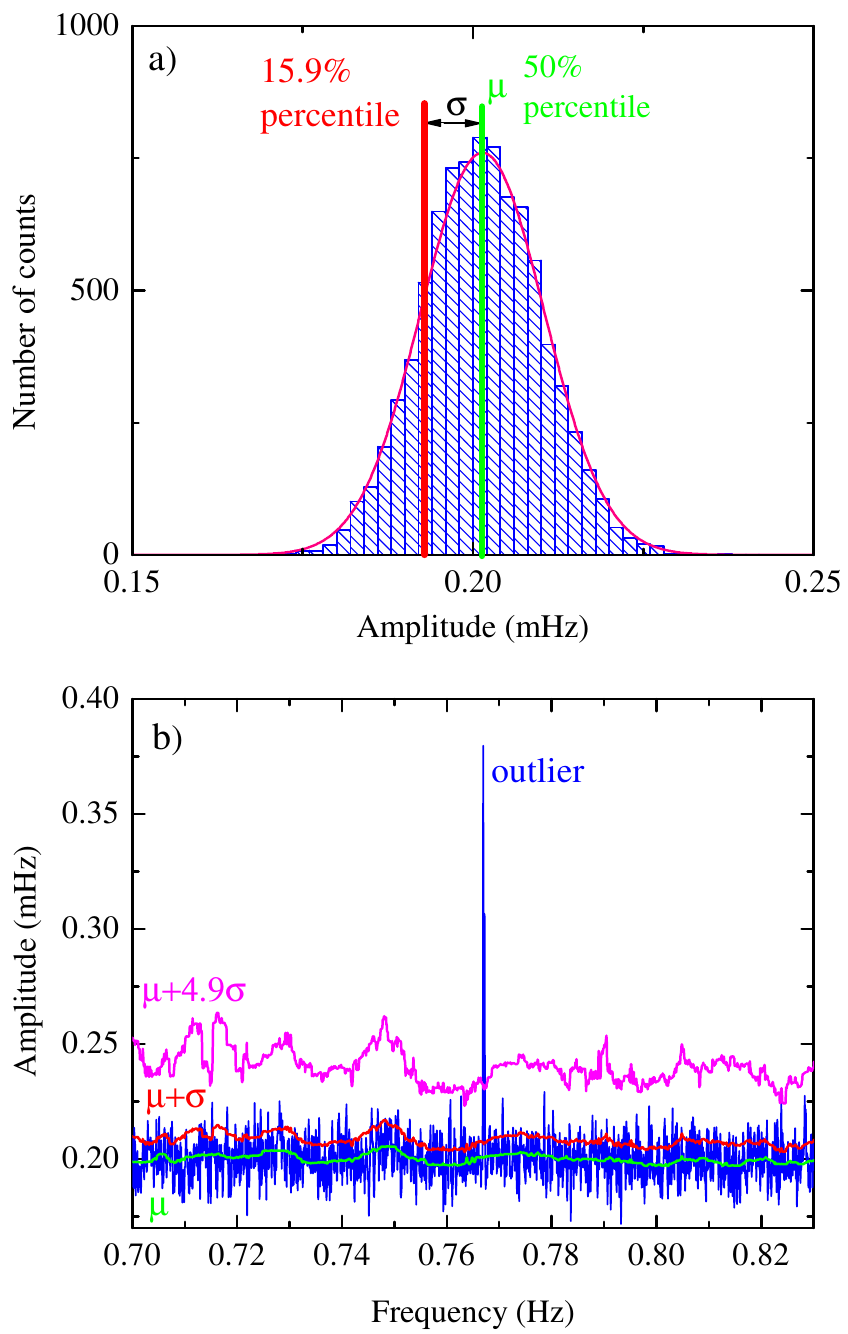}
    \caption{ a) Distribution of noise 
    calculated from data acquired in Experiment 1, in a frequency window 0.5$-$1.1\,Hz. The noise distribution is well approximated with a Gaussian. Here, $\mu$ is the median, and $\sigma$ is the standard deviation. 
    b) An example of a spurious peak observed in the frequency window 0.7$-$0.83\,Hz in Experiment 1, denoted as an outlier. 
} 
    \label{fig:Histogram}
\end{figure}

An amplitude spectrum is obtained by taking the square root of the final power spectrum obtained as described in Sec.\,\ref{"subsubsec:SI-Data analysis-Averaged power spectra"}.  The noise of this spectrum (with the exclusion of frequency bins containing obvious spurious signals) follows a Gaussian  distribution, as shown in Fig.\,\ref{fig:Histogram}\,a. To identify spurious signals within the amplitude spectrum, the amplitude of each frequency bin is compared with the noise in its vicinity. 

The standard deviation $\sigma$ of the noise gives a natural scale to set a threshold to discriminate the spurious signals from the noise background. An example of such a signal, denoted as outlier, is shown in Fig.\ref{fig:Histogram}\,b. 

To obtain the $\sigma$ of the amplitude spectrum, its baseline is first removed. This is done to improve the accuracy in the subsequent calculation of the moving  $\sigma$ of the spectrum, primarily because the varying noise power at low frequencies (the pink or 1/\textit{f} noise) impacts the calculation. To compute this baseline, a moving 50\%-percentile filter is applied to the spectrum within a frequency window of 100 bins. (A 50\%-percentile filter applied to normally distributed data, provides the mean value of the data set; see Fig.\,\ref{fig:Histogram}a). After subtracting the computed baseline from the spectrum, 
a moving  50\%-percentile filter and a moving  15.9\%-percentile filter are applied to the spectrum, using the same window width (i.e.\,100 bins). (A 15.9\%-percentile filter applied to a set of normally distributed data gives the value that is 1\,$\sigma$ below the mean value; see Fig.\,\ref{fig:Histogram}a.) Subtracting the respective filter outputs yields the moving $\sigma$ of the amplitude spectrum. 

However, there are flaws in applying a percentile filter in the low-frequency end of the spectra. First, the application of the filter within a window of $ N$ bins, naturally fails for the first $N/2$ bins that need to be discarded. 
Second, a successive application of the filter to compute the spectrum's $\sigma$ would force us to discard an additional $N/2$ bins, for a total of $N$ bins. Third, the rapidly varying $1/f$ noise impacts the precision of the calculation. 

To 
compute the spectrum baseline at low frequencies and avoid discarding data points, we instead compute the  baseline in the low-end by fitting to the noise (as mentioned, dominated by the $1/f$ contribution). The fit is to the first 200 bins of the spectrum (i.e. the range up to 14\,mHz). After the subtraction of this baseline, we then proceed with calculating $\sigma$, by applying a moving 50\%-percentile filter and a moving 15.9\%-percentile filter to the spectrum (within a reduced-size window of 70 bins below 7\,mHz), as explained above.

However, it is not possible to reliably calculate the moving $\sigma$ at the lowest frequencies. Although, in principle, we could probe for oscillations at the lowest frequency  $70\,\rm{\mu}$Hz (i.e. the width of a bin), there are two issues: first, the need for sizable width of the window used in the applied percentile filters (windows that are tens of bins wide are needed); second (as mentioned above), the increasing $1/f$ noise at low frequencies. Being unable to reliably compute moving $\sigma$ below 1\,mHz, we conservatively employ computed  $\sigma$ for frequencies $f_{\rm{C}}\geq 2.5$\,mHz.   


From the computed $\sigma$ spectra, a threshold for investigating spurious signals in the $\delta f/f$ spectra is set considering the `look-elsewhere`  effect\,\cite{Scargle1982}, and the respective bin numbers in the spectra of Experiment 1 ($n=188364$) and Experiment 2 ($n=2879971$).  The computed  95$\%$\,C.L. thresholds are, respectively, set to 4.9\,$\sigma$ and 5.32\,$\sigma$. 

\begin{figure}[t]
    \centering
    \includegraphics[width=\columnwidth]{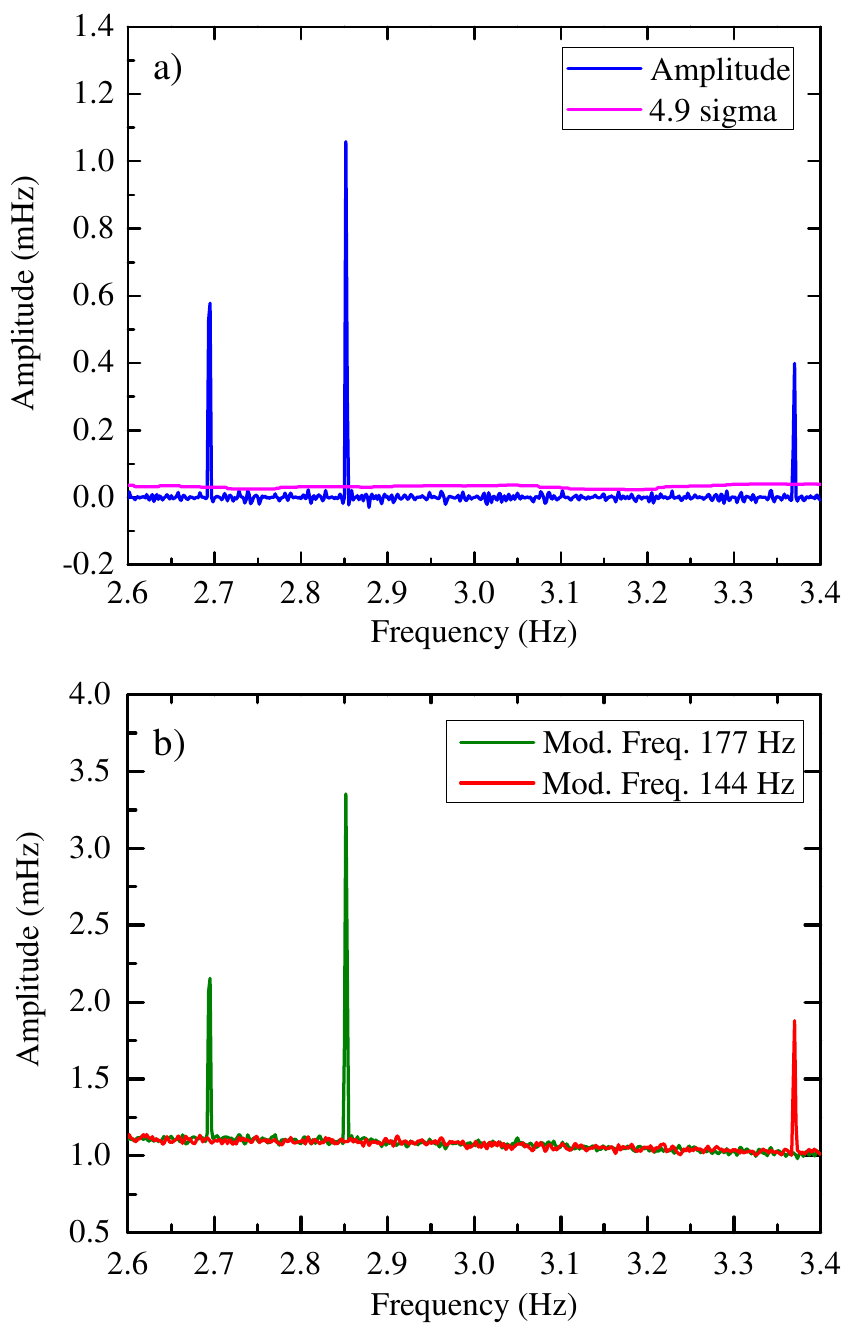}
    \caption{ 
    a) Identification of candidate peaks. Three peaks are shown in the frequency window 2.6$-$3.4\,Hz for the low-frequency measurement in Experiment 1. b) Intercomparison of spurious peaks in separate amplitude spectra averaged by the data acquired with different modulation frequencies. All the peaks shown here are detected only in one of the spectra for the respective frequency bin and can be 
    eliminated. 
} 
    \label{fig:Spurious}
\end{figure}

\subsubsection{Investigating spurious signals}

Peaks in the obtained amplitude spectra with size exceeding the set threshold are 
potential UBDM signals  (see Fig\,\ref{fig:Spurious}\,a). To investigate whether these peaks are real or spurious, various measures were taken, which are discussed separately for Experiments 1 and 2.  

\emph{Experiment 1---} A total of ten candidate peaks 
were identified in the amplitude spectrum of the low-frequency run of Experiment 1. An example of a strategy to investigate them is shown in Fig.\,\ref{fig:Spurious}. The amplitude spectra separately averaged for the different modulation frequencies are compared, as shown in Fig.\,\ref{fig:Spurious}\,b. If the 
candidate peaks are not observed 
at the same frequency in all amplitude spectra, they can be eliminated directly. In this way, seven out of the ten peaks of the low-frequency run can be 
excluded from being UBDM candidates. Moreover, in the respective frequency bins of the eliminated peaks, the detection threshold has to be recalculated, because data segments used to obtain the averaged amplitude spectra that include the spurious peaks have to be abandoned. For the remaining three 
candidate peaks appearing in both averaged amplitude spectra at the same frequency, data from the high-frequency run are utilized additionally for intercomparison.

Investigation of spurious peaks in the amplitude spectrum of the high-frequency run is done similarly. Of the 231 peaks observed, 216 were eliminated via intercomparison among amplitude spectra acquired for the three different modulation frequencies used in the run. To further check the remaining 15 peaks, additional data were taken and compared with the main set, as follows: i) with another (fourth) modulation frequency,  resulting in additional elimination of seven  peaks; ii) using a different signal generator (see Fig.\,\ref{fig:Rb_setup}), that cleared out the remaining eight peaks.

\emph{Experiment 2---}The checks of spurious peaks in experiment 2 are done similarly. A total of 983 peaks are detected in the averaged amplitude spectrum, 467 of which can be eliminated via intercomparison among amplitude spectra acquired for the three modulation frequencies of the run.  In addition, 514 peaks are excluded by comparing the main data set with an additional set re-sampled using a fourth modulation frequency. Two peaks remain after this process, that are observed at frequencies 34\,Hz and 68\,Hz in all amplitude spectra. These are excluded by acquiring additional data sets with a different function generator (see Fig.\,\ref{fig:Dy_setup}a), as it was done in the high-frequency run of Experiment 1. 

\subsection{Calibrations}
\begin{figure}[t]
    \centering
    \includegraphics[width=\columnwidth]{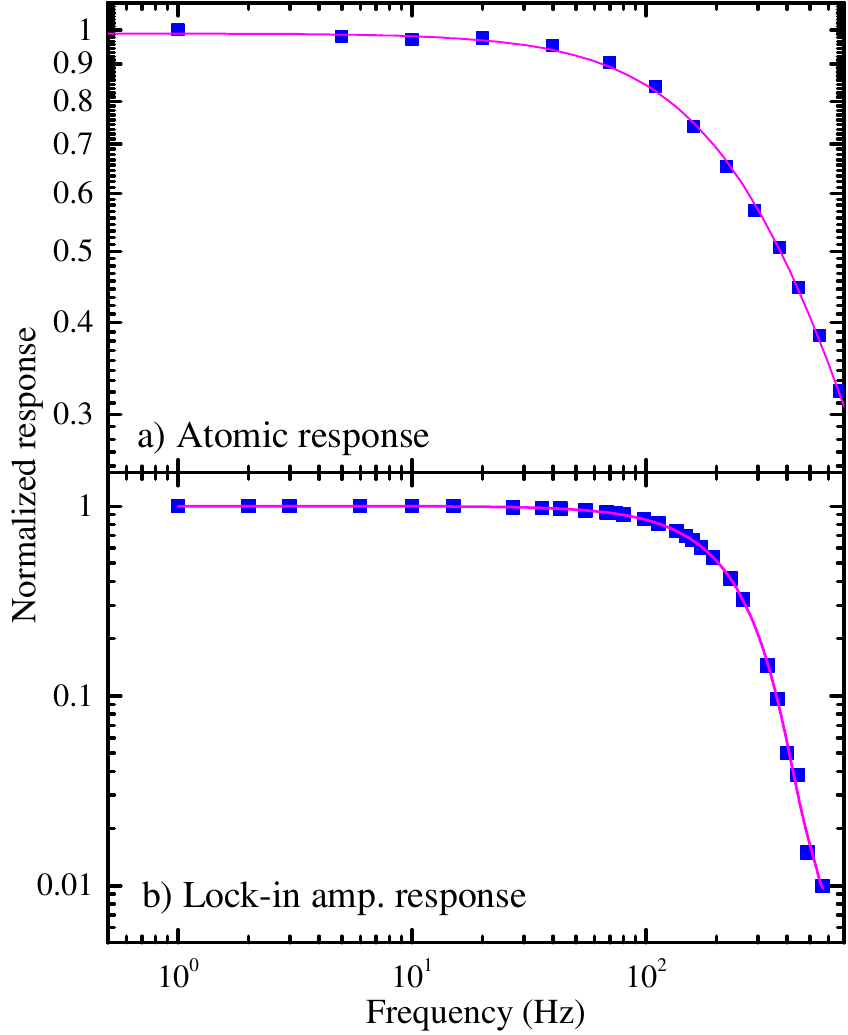}
    \caption{ a) Normalized atomic calibration.  b) Normalized lock-in amplifier calibration. The magenta lines represent a fit to measured data points.} 
    \label{fig:Atomicandlockincalibrations}
\end{figure}

\emph{Atomic calibration ---} Calibration of the atomic response in Experiment 1 is done for the parameters of the high-frequency experimental run. The rf frequency is tuned on the side of the hyperfine resonance, and frequency modulation with a fixed amplitude is imposed on the rf field 
applied to the atoms. The amplitude of the induced oscillation in the light transmitted through the Rb vapor (i.e. the probe of the hyperfine resonance) is recorded with varying modulation frequency (Fig.\,\ref{fig:Atomicandlockincalibrations}\,a). A curve fitted to these calibration data is applied to the $\delta f/f$ spectrum of Experiment 1 to account for the slight reduction in the atomic response in the high-frequency end of our 2.5\,mHz$-$200\,Hz search for oscillating effects. 

In Experiment 2, the response of Dy atoms is considered uniform within the 2.5\,mHz$-$200\,Hz investigated range, since the observed 50-kHz linewidth of the Dy transition (of natural linewidth $\approx$20\,kHz, that is broadened due to the finite transit time of atoms through the interaction region and due to the
power of the applied rf field) 
is much greater than the frequencies our search covers. 

\emph{Lock-in amplifier calibration---} Calibration of the lock-in-amplifier response is necessary for both Experiment 1 and 2 to account for low-pass filtering of the amplifier output that becomes significant at the higher frequencies of our search. The calibration is done with the same settings as those in the actual experiments (a 1-ms time constant and a filter slope of 18\,dB/octave). A sinusoidal signal with a 10-kHz frequency is measured with the lock-in amplifier. The signal amplitude is modulated slightly at a frequency $f_{\rm {mod}}$, causing the amplifier output to oscillate at the same frequency. (This oscillation is the signature expected in the presence of UBDM-induced oscillations, thus the calibration setup emulates the actual experiments.) The amplitude of the lock-in-output oscillation is measured against $f_{\rm {mod}}$; the result is shown in Fig.\,\ref{fig:Atomicandlockincalibrations}\,b. We see that the response is attenuated by $\approx 2 \times$ for the highest frequency within our search (200\,Hz). We apply corrections to all the UBDM  constraints to account for the lock-in-amplifier response.


\subsection{Artificial signal injections}

To check the validity of the set  95$\%$ C.L. detection thresholds in the $\delta f/f$ spectra, we  inject artificial signals of known amplitudes, frequencies and phases to the experimentally obtained 
time-series data. This is done for 20 frequency values spanning our 2.5\,mHz$-$200\,Hz UBDM-search range,  and several phase values for each frequency. The amplitudes of these injected signals approximately match the previously set detection thresholds. We find that the signals appear in the resulting $\delta f/f$ spectra with amplitudes $\approx$1.3 times smaller than expected. We multiply the $\delta f/f$ spectra and the obtained detection thresholds by this factor, to   re-calibrate apparatus sensitivity. 
The observed discrepancy is the result of computing the injected-signal amplitudes using amplitude spectra. If one instead uses power spectra, the computed amplitudes are as expected. 
The discrepancy occurs in the limit of small injected signals (i.e. signals with amplitudes on the order of the noise level). In the limit of large signals (i.e. amplitudes that are at least tens of times greater than the noise level)
the obtained amplitudes in the  $\delta f/f$ spectra tend to the expected size, thus providing a check for the apparatus calibration. 

\subsection{Dependence of  $m_{\rm{p}}$ and $g_{\rm{nuc}}$ on the field $\phi(t)$}
Let us consider QCD axion models where a pseudo-scalar field,  $\phi(t)$, the axion, couples to the gluon field of strength $G^{\mu\nu}$, contributing a term to the Lagrangian density:
\begin{equation}
\mathcal{L} \supset\frac{g_s^2}{32\pi^2}\frac{\phi}{f_{\rm{\phi}}} G^{\mu\nu}\widetilde{G}_{\mu\nu}\,,
\label{eq:axion_gluon}
\end{equation}
where $f_{\phi}$ is the axion decay constant, $g_s$ is the strong coupling constant and $\widetilde G_{\mu\nu}$ is the dual gluon field strength. We keep the color indices implicit. 
Considering interactions at energies much lower than the QCD confinement scale,
$\Lambda_{\rm QCD}$, this term gives rise to 
coupling of the axion to the hadrons. Specifically, the pion mass depends on the axion field as \cite{Ubaldi:2008nf}, 
\begin{equation}
\!\!\!
m_{\pi}^2(\theta_{\rm eff}) = \frac{\Lambda_{\rm QCD}^3}{f_{\pi}^2} \sqrt{m_u^2+m_d^2+2 m_u m_d \cos(\theta_{\rm eff})}\,,
\label{eq:m_pion}
\end{equation}
where we have defined $\theta_{\rm eff}=\bar\theta+\phi/f_{\phi}$ with $\bar\theta = \theta+{\rm arg\,det\,}(M)$, $\theta$ being the QCD $\theta$ angle and $M$ is the quark-mass matrix. The parameters $m_u$ and $m_d$ denote the masses of the up and down quarks, respectively, while $f_\pi\simeq 92\, {\rm MeV}$ is the pion decay constant. 
The potential of the axion can be written as $V(\theta_{\rm eff})=-m_\pi^2(\theta_{\rm eff})f_\pi^2$~\cite{DiVecchia:1980yfw} and is minimized when $\left<\theta_{\rm eff}\right>=0$. 
From this follows $(\left<\phi\right>/f_{\phi})_{\rm min}=-\bar\theta$ 
~\cite{Vafa:1984xg}.
Thus, by relaxing to the CP conserving vacuum, the axion solves the strong-CP problem dynamically~\cite{Peccei:1977ur,Peccei:1977hh,Weinberg:1977ma,Wilczek:1977pj,Kim:1979if,Shifman:1979if,Zhitnitsky:1980tq,Dine:1981rt}. 
An axion field coherently oscillating around its minimum 
may account for DM in the present universe~\cite{Preskill:1982cy,Abbott:1982af,Dine:1982ah} and can be represented as $\theta_{\rm eff}(t)= (\phi_0/f_{\phi})\cos(m_{\phi} t)$ where $\phi_0$ is the field amplitude and $m_{\phi}$ is the axion mass.

To see how the QCD axion induces time variation of the FCs at the quadratic order, we expand the pion mass close to the minimum of the QCD axion potential and obtain  
\begin{eqnarray}
\frac{\delta\,m_{\pi}^2}{m_{\pi}^2} &=& \frac{m_{\pi}^2(\theta_{\rm eff})-m_{\pi}^2(0)}{m_\pi^2(0)}\\ 
&\simeq &  - \frac{m_u m_d\,\theta_{\rm eff}^2(t)}{2(m_u+m_d)^2} \simeq - 0.11\frac{\phi(t)^2}{f_{\phi}^2}\,,
\label{eq:mpi_theta}
\end{eqnarray}

\noindent where we used the values of $m_{u,d}$ from~\cite{Workman:2022ynf}. 
As shown in~\cite{Kim:2022ype}, time variation of the pion mass leads to time variation of the proton mass, $m_{\rm{p}}$, as 
\begin{equation}
\frac{\partial \ln m_{\rm p}}{\partial \ln m_\pi^2} = 0.06\,.
\label{eq:mn_mpi}
\end{equation}
The nuclear $g$-factor, $g_{\rm nuc}$, also depends on the pion mass; for ${}^{87}$Rb we have \cite{Kim:2022ype} 

\begin{equation}
\frac{\partial \ln g_{\rm nuc}}{\partial \ln m_\pi^2} = -0.024\,.
\label{eq:g_nuc}
\end{equation}
\noindent From Eqs.\, \eqref{eq:mpi_theta},\eqref{eq:mn_mpi}, and \eqref{eq:g_nuc} one obtains the dependence of $m_{\rm{p}}$ and $g_{\rm{nuc}}$ on the field $\phi(t)$ shown in Eq.\,(4) and Eq.\,(5) of the main text.

\end{document}